
\documentstyle[12pt]{article}
\newcommand{\be}{\begin{equation}}
\newcommand{\ee}{\end{equation}}

\begin{document}
\begin{center}
{\large {\bf On a Lie algebraic approach of quasi-exactly solvable potentials with two known eigenstates}}\\
\vspace{0.5in}
Y.BRIHAYE\footnote{email: Yves.Brihaye@umh.ac.be}, \\
{\it Department of Mathematical Physics, 
University of Mons-Hainaut, Place du Parc, B-7000 MONS (Belgium)} \\
 N. DEBERGH\footnote{Chercheur, Institut Interuniversitaire des
Sciences Nucl\'eaires, Bruxelles, email: Nathalie.Debergh@ulg.ac.be} \\
{\it Theoretical Physics,
Institute of Physics (B5),
University of Li\`ege,
B-4000 LIEGE  (Belgium)}\\
and J. NDIMUBANDI\footnote{email: jndimubandi@yahoo.fr}\\
{\it University of Burundi, Department of Mathematics, P.O. Box 2700, BUJUMBURA (Burundi)} \\
\vspace{0.2in}
\end{center}
\vspace{0.5in}
\date{\today}

\begin{abstract}
We compare two recent approaches of quasi-exactly solvable Schr\" odinger equations, the first one being related to finite-dimensional representations of $sl(2,R)$ while the second one is based on supersymmetric developments. Our results are then illustra
ted on the Razavy potential, the sextic oscillator and a scalar field model.
\end{abstract}
\newpage

\section{Introduction}
Quasi-exactly solvable (Q.E.S.) equations are those for which a finite number of analytic solutions can be determined. The first examples of such equations having appeared in the literature are one-dimensional Schr\" odinger ones connected with the Razavy
 potential [1] as well as the sextic oscillator [2]. Since that first step, Q.E.S. Schr\" odinger equations have been listed [3] according to their relation with the finite-dimensional representations of the Lie algebra $sl(2,R)$. 
Indeed, generically, a QES Schr\" odinger equation
\be
(-\frac{1}{2} \frac{d^2}{dx^2} + V(x))\psi(x) = E \psi(x)
\ee
can be written as
\be
(p_4(z)\frac{d^2}{dz^2}+p_3(z)\frac{d}{dz}+p_2(z))\phi(z) = E \phi(z)
\ee
through ad-hoc changes of variables and functions [3]
\be
z=z(x) \; , \; \phi = \exp(\chi )\psi
\ee
if $p_a(z)$ ($a=2,3,4$) refer to polynomials of order $a$ in $z$. The differential operator in (2) can be expressed as a quadratic 
combination (including the linear terms)  of the operators
\be
j_+ = -z^2\frac{d}{dz}+Nz,
\ee
\be
j_0=z\frac{d}{dz}-\frac{N}{2},
\ee
\be
j_-=\frac{d}{dz}, \; \; \; N=0,1,2,...
\ee
The operators (4)-(6) actually generate the $sl(2,R)$ algebra through the commutation relations
\be
[j_0 , j_{\pm}]=\pm j_{\pm} \; , \; [j_+ , j_-]=2j_0.
\ee
The crucial point which reveals that Eq. (2) is indeed Q.E.S. is the introduction of the nonnegative integer $N$ in (4)-(5). Indeed, this ensures the operators (4)-(6) to preserve the space of polynomials of order $N$
\be
P(N)=\{1,z,z^2,...,z^N\}
\ee
and so does the operator in (2). 
Searching for the eigenvalues of (2) 
(or equivalently of (1)) restricted to 
the $(N+1)$-dimensional space $P(N)$
reduces the problem to an algebraic one; in other words starting from a differential equation, we are led to an algebraic one giving the $(N+1)$ analytic solutions of (2) (or (1)).
\par
Recently, the classification of  Schr\" odinger  equations admitting
at least two algebraic eigenvalues has been addressed in
[4] through supersymmetric techniques. More precisely, it has been proved that if $V(x)$ looks like a supersymmetric partner i.e. if
\be
V(x) = \frac{1}{2} W^2(x) - \frac{1}{2} \frac{dW(x)}{dx}
\ee
then (1) is Q.E.S. in the sense that {\it two} of its eigenstates can be found in an analytical way. They are given by
\be
\psi_0(x) = e^{-\int{W(x)dx}} \; \; , \; E_0=0
\ee
and
\be
\psi_1(x) = (-\frac{d}{dx}+W(x)) e^{-\int{W_1(x)dx}} \; \; , \; E_1=\epsilon
\ee
provided the functions $W(x)$ and $W_1(x)$ are such that
\be
W(x)=\frac{1}{2}W_+(x) + \frac{\epsilon}{W_+(x)} - \frac{1}{2W_+(x)}\frac{dW_+(x)}{dx},
\ee
\be
W_1(x)=\frac{1}{2}W_+(x) - \frac{\epsilon}{W_+(x)} + \frac{1}{2W_+(x)}\frac{dW_+(x)}{dx}
\ee
whatever $W_+(x)$ is, as far as both $W(x)$ and $W_1(x)$ are free of singularities.
\par
In the present paper, we propose to revisit these supersymmetric developments under the Lie algebraic point of view based on $sl(2,R)$ and its representation space (8)
considered for $N=1$. This will lead to a possible comparison between the Turbiner approach and the Tkachuk one. We also illustrate our results on three examples connected to the Razavy potential [1], the sextic oscillator [4] and a toy model in scalar fi
eld theory. Finally, some conclusions are presented.
\section{The Lie algebra $sl(2,R)$ inside the $N=1$-context}
Let us first rewrite (9)-(11) in terms of the function $W_+(x)$. We successively have
\be
V(x) = \frac{1}{8}W_+^2 + \frac{1}{2}\frac{\epsilon^2}{W_+^2}+\frac{\epsilon}{2}-\frac{1}{2}\frac{dW_+}{dx}-\frac{1}{8W_+^2}(\frac{dW_+}{dx})^2+\frac{1}{4W_+}\frac{d^2W_+(x)}{dx^2},
\ee
\be
\psi_0(x)=\sqrt{W_+}e^{-\frac{1}{2}\int{W_+(x)dx}}e^{-\epsilon \int{\frac{dx}{W_+}}},
\ee
\be
\psi_1(x)=\sqrt{W_+}e^{-\frac{1}{2}\int{W_+(x)dx}}e^{\epsilon \int{\frac{dx}{W_+}}}.
\ee
In order to satisfy the Turbiner characteristics that the differential operators act on $P(1)$, we propose to define $\phi(x)$ (see (3)) as
\be
\psi_n(x) =\sqrt{W_+}e^{-\frac{1}{2}\int{W_+(x)dx}}e^{-\epsilon \int{\frac{dx}{W_+}}}\phi_n(x) ,\; \; n=0,1
\ee
and to consider the following change of variables
\be
z = e^{2\epsilon \int{\frac{dx}{W_+}}}.
\ee
It is then straightforward to be convinced that
\be
\phi_0(z)=1, \; \phi_1(z)=z
\ee
and that the Schr\" odinger equation (1) with $V(x)$=(14) is now simply
\be
(-2\epsilon^2\frac{z^2}{W_+^2}\frac{d^2}{dz^2}+\epsilon z\frac{d}{dz})\phi_n(z) \equiv T \phi_n(z) = E_n \phi_n(z)
\ee
with $E_0=0,E_1=\epsilon$.Compared with (2), this equation implies that either the Turbiner and Tkachuk approaches are equivalent and this is ensured iff $W_+$ can be put on the form
\be
W_+(z) = \pm \frac{z}{\sqrt{c_0+c_1z+c_2z^2+c_3z^3+c_4z^4}}
\ee
where $c_j \; (j=0,1,...,4)$ are arbitrary real numbers {\it or} these approaches differ ($W_+(z) \neq $ (21) in this case) in the sense that the Tkachuk operator is the Turbiner one supplemented by an element of the kernel.
\par
Let us now turn to an illustration of these results on three examples.
\section{Examples}
\subsection{Example 1}
The Razavy potential [1] is associated [4] with the following
form for  $W_+(x)$~:
\be
W_+(x)=A \sinh(\alpha x)
\ee
with $A$, $\alpha > 0$ and the non-vanishing energy is $\epsilon = \frac{\alpha A}{2}$. This leads, through (18), to the  new variable
\be
z = \tanh (\frac{\alpha}{2}x)
\ee
and the corresponding $W_+$
\be
W_+(z)= 2A \frac{z}{1-z^2}.
\ee
It thus coincides with (21) and the two approaches are equivalent. The algebra $sl(2,R)$ generated by (4)-(6) with $N=1$ is then the one subtending such an example and it is immediate to see that
\be
T=-\frac{\alpha^2}{8}j_+^2+(\frac{\alpha A}{3}-\frac{\alpha^2}{12})j_+j_-+(\frac{\alpha A}{3}+\frac{\alpha^2}{6})j_0^2-\frac{\alpha^2}{8}j_-^2+(\frac{\alpha A}{6}+\frac{\alpha^2}{12})j_0.
\ee
\subsection{Example 2}
The sextic oscillator considered in [4] corresponds to the choice
\be
W_+(x) = ax+bx^3
\ee
with $a,b>0$ and $\epsilon=\frac{a}{2}$. The variable $z$ now reads
\be
z=\frac{x}{\sqrt{a+bx^2}}
\ee
while $W_+$ takes the form
\be
W_+(z) = \frac{a^{\frac{3}{2}}z}{\sqrt{(1-bz^2)^3}}.
\ee
Clearly, it does not satisfy the requirement (21). The generators (4)-(6) with $N=1$ are thus unsufficient in order to explain the Q.E.S. features of this sextic oscillator. More precisely, we have
\be
T = -\frac{3b^2}{2a}j_+^2 + (\frac{a}{3}-\frac{b}{2a})j_+j_-+(\frac{a}{3}+\frac{b}{a})j_0^2-\frac{1}{2a}j_-^2+(\frac{a}{6}+\frac{b}{2a})j_0+\frac{b^3}{2a}z^6\frac{d^2}{dz^2}.
\ee
The presence of a non-vanishing element of the kernel is thus necessary in this case.
\subsection{Example 3}
Another type of physical applications in which equations like (2) come out is in the study of the stability of solitons in low-dimensional scalar field theory. Classical solutions of the soliton type [5] are in general difficult to study analytically in r
ealistic field theories; therefore their study in toy model is often fruitfull and interesting. In this respect, the scalar fields theories with polynomial interaction constitute excellent toy models [6] and classical solutions in such models were analyze
d in many papers, see e.g. [7]-[8].
\par
Let us thus consider the field theory of a scalar field $\rho(t,y)$ (in 1+1 dimensions) 
self-interacting through a scalar potential, say $V_s(\rho)$
\be
{\cal L} =\frac{1}{2}(\frac{\partial \rho}{\partial t})^2-\frac{1}{2}(\frac{\partial \rho}{\partial y})^2-V_s(\rho).
\ee
The static classical equation reads
\be
\frac{\partial^2 \rho}{\partial y^2}=\frac{d}{d\rho}V_s(\rho)
\ee
and admits a first integral
\be
\frac{1}{2}(\frac{\partial \rho}{\partial y})^2-V_s(\rho)=-\frac{K}{2}.
\ee
The constant $K$ determines the boundary (or periodicity) condition of the solution. Given a static solution and its value of $K$, say $\bar \rho(y), \bar K$, the stability of the solution can be studied by solving the Schr\" odinger equation
\be
(-\frac{d^2}{dy^2}+\frac{d^2V_s(\bar \rho)}{d \rho^2})\eta(y) = \omega^2 \eta(y).
\ee
where $\eta(y)$ describes a fluctuation about the soliton (i.e. $\rho(y) = \bar \rho(y) + \eta(y)$).

If $V_s$ is chosen according to
\be
2V(\rho)-K=(\rho^2-A)(\rho^2-B)(\rho^2-C)
\ee
and using the square of the soliton profile as a new variable,
\be
u = \bar \rho^2(y),
\ee
Eq. (33) can be written explicitely even in the absence of an explicit knowledge
of the soliton function $\bar \rho(y)$; it reads
\begin{eqnarray}
&&(4u(u-A)(u-B)(u-C)\frac{d^2}{du^2}+2(4u^3-3u^2(A+B+C)\nonumber \\
&&+2u(AB+AC+BC)-ABC)\frac{d}{du}-15u^2+6u(A+B+C))\eta(u)\nonumber \\
&&=((AB+AC+BC)-\omega^2)\eta(u).
\end{eqnarray}
Despite of the fact that it is of the form (2), it does not preserve any space $P(N)$. Nevertheless it admits one solution given by
\be
\eta_0(u) = \sqrt{(u-A)(u-B)(u-C)} \ \ , \ \  \omega_0^2=0
\ee
which corresponds to the zero mode associated with the invariance of the theory
under translations.  If $A=2(B+C)$, then a second explicit
solution does exist which is given by
\be
\eta_1(u) = \sqrt{(u-A)}(u-\frac{B+C}{2}) \ \ , \ \  \omega_1^2=2(B-C)^2
\ee
From now on we restrict ourselves to the case $A=2(B+C)$, we redefine
\be
\eta(u)=\sqrt{(u-A)(u-B)(u-C)} \phi(u)
\ee
and we perform the change of variable
\be
z=\frac{1}{\sqrt{(u-B)(u-C)}}(u-\frac{B+C}{2}).
\ee
Equation (36) then becomes
\begin{eqnarray}
&&((3(B+C)^2(z^2-1)^2+2(B-C)(B+C)z(z^2-1)^{\frac{3}{2}}-(B-C)^2z^2(z^2-1))\frac{d^2}{dz^2}\nonumber \\
&&+2(B-C)^2z\frac{d}{dz}-\omega^2)\phi(z)=0.
\end{eqnarray}
Clearly, it looks like (20) with $\epsilon=2(B-C)^2$ and, consequently, it is consistent with
Tkachuk's approach.  The formalism of Turbiner being recovered 
only for $B=-C$, in this case the relevant operator reads
\begin{eqnarray}
&&T=-4B^2z^2(z^2-1)\frac{d^2}{dz^2}+8B^2z\frac{d}{dz} \nonumber \\
&& \; \; =-4B^2 j_+^2+4B^2j_0^2+8B^2j_0+3B^2.
\end{eqnarray}

Coming back to the formalism of \cite{4}, a direct computation of the function
$W_+$ leads to
\be
W_+=\frac{4\sqrt{2}B}{\sqrt{z^2-1}}
\ee
and using the change of variables (18) i.e.
\be
x=-\frac{1}{2\sqrt{2}B}\arctan(\frac{1}{\sqrt{z^2-1}})
\ee
we obtain
\be
W_+=4\sqrt{2}B \tan (-2\sqrt{2}Bx).
\ee
The Schr\"odinger equation (33) 
then acquires a supersymmetric form with 
superpotential and potential given respectively by
\be
W(x)=2\sqrt{2}B \cot (-2\sqrt{2}Bx)+ 3\sqrt{2}B \tan (-2\sqrt{2}Bx)
\ee
and
\be
V(x)=15B^2 \tan^2(-2\sqrt{2}Bx)+14B^2.
\ee
These are particular  P\" oschl-Teller (super)potentials [9].

We further studied the 
supersymmetric partner equation, i.e. with a potential given by
\be
\bar V(x) = \frac{1}{2}W^2(x)+\frac{1}{2}\frac{dW(x)}{dx} \ \ ,
\ee
Again, it corresponds  to a P\" oschl-Teller potential~:
\be
\bar V(x) = 8B^2 \cot^2(-2\sqrt{2}Bx)+3B^2 \tan^2(-2\sqrt{2}Bx)+10B^2.
\ee
As is well known from supersymmetric considerations, a solution of the Schr\" odinger equation
\be
(-\frac{1}{2}\frac{d^2}{dx^2}+\bar V(x))\bar \psi(x)=E \bar \psi(x)
\ee
is automatically known. It reads
\be
\bar \psi_0(x)=e^{-\int{W_1(x)dx}}, E_0=8B^2
\ee
which with (13) and (45) gives rise to
\be
\bar \psi_0(x)=(\cos(-2\sqrt{2}Bx))^{-\frac{1}{2}}(\sin(-2\sqrt{2}Bx))^{-1}, E_0=8B^2.
\ee
Compared to the known solutions [9]
\begin{eqnarray}
&&\bar\psi_{n+1}(x)=(\cos(-2\sqrt{2}Bx))^{\frac{3}{2}}(\sin(-2\sqrt{2}Bx))^{2}\nonumber \\
&&_2F_1(-n,7/2+n;5/2;\sin^2(-2\sqrt{2}Bx)), \nonumber \\
&&E_{n+1}=B^2(7+4n)^2, n=0,1,2,...
\end{eqnarray}
The function (52) formally constitutes a 
supplementary solution although only on a   mathematical level
since it goes to infinity with the potential $\bar V(x)$. Anyway the solutions $\bar\psi_{n+1}(x)$ given in (53) also lead to new solutions of the original Schr\" odinger equation (1) with (47)
\begin{eqnarray}
&&\psi_{n+2}(x)=15(\cos(-2\sqrt{2}Bx))^{\frac{5}{2}}(\sin(-2\sqrt{2}Bx))\nonumber \\
&&_2F_1(-n,7/2+n;5/2;\sin^2(-2\sqrt{2}Bx))\nonumber \\
&&-4n(7/2+n)(cos(-2\sqrt{2}Bx))^{\frac{5}{2}}(\sin(-2\sqrt{2}Bx))^3\nonumber \\
&&_2F_1(-n+1,9/2+n;7/2;\sin^2(-2\sqrt{2}Bx))
\end{eqnarray}
with the same energies $E_{n+1}$. Coming back to the original scalar field model and thus to the variable $u$, these new solutions take the form
\begin{eqnarray}
&&\eta_{n+2}(u)=15B^6u^{-\frac{7}{2}}\sqrt{u^2-B^2}_2F_1(-n,7/2+n;5/2;\frac{u^2-B^2}{u^2}) \nonumber \\
&&-4n(7/2+n)u^{\frac{1}{2}}(\sqrt{u^2-B^2})^3\; \; _2F_1(-n+1,9/2+n;7/2;\frac{u^2-B^2}{u^2}).
\end{eqnarray}

\section{Conclusions}
We have put in evidence the connection between two approaches of Q.E.S. equations with two known eigenstates: the Turbiner one [2] is based on the two-dimensional representation of $sl(2,R)$ while the Tkachuk one [4] deals with supersymmetric quantum mech
anics. We have shown that both approaches actually are equivalent ones up to the eventual presence of an element of the kernel. We also have exploited such a connection and the supersymmetric characteristics in order to produce new solutions of a toy mode
l from scalar field theory.\\

{\bf ACKNOWLEDGMENTS}
\par
One of us (J.N.) would like to thank the Theoretical Physics Group of the University of Liege for its warm hospitality during his stay in Belgium and the "Agence Universitaire de la Francophonie" for its financial support.

\end{document}